%% file: sample-sigconf.tex
\documentclass[sigconf,9pt,prologue,table]{acmart}

\definecolor{benchGreen}{RGB}{133, 182, 114}
\definecolor{benchRed}{RGB}{230, 141, 143}
\definecolor{stargraphMillion}{RGB}{0, 143, 251}
\definecolor{stargraphSnabbdom}{RGB}{1, 227, 150}
\definecolor{stargraphSimpleVirtualDom}{RGB}{254, 176, 24}
\definecolor{stargraphHundred}{RGB}{255, 69, 95}
\definecolor{stargraphVirtualDom}{RGB}{119, 93, 208}

\copyrightyear{2023}
\acmYear{2023}
\setcopyright{acmlicensed}\acmConference[SAC '23]{The 37th ACM/SIGAPP Symposium on Applied Computing}{March 27-March 31, 2023}{Tallinn, Estonia}
\acmBooktitle{The 37th ACM/SIGAPP Symposium on Applied Computing (SAC '23), March 27-March 31, 2023, Tallinn, Estonia}
\acmPrice{15.00}
\acmDOI{10.1145/3555776.3577683}
\acmISBN{978-1-4503-9517-5/23/03}
\acmArticle{4}

\usepackage{booktabs} 
\usepackage{color}
\usepackage{colortbl}
\usepackage{booktabs}
\usepackage{textcomp}
\usepackage{csquotes}
\usepackage[normalem]{ulem}
\usepackage{array}
\usepackage{graphicx}

\begin{document}

\title{Million.js: A Fast Compiler-Augmented Virtual DOM for the Web}
  
\renewcommand{\shorttitle}{Million.js}

\author{Aiden Bai}
\affiliation{%
  \institution{Camas High School}
  \city{Camas}
  \state{Washington}
  \country{USA}
}
\orcid{0000-0002-3676-3726}
\email{aiden.bai05@gmail.com}

\renewcommand{\shortauthors}{A. Bai}

\begin{abstract}
Interactive web applications created with declarative JavaScript User Interface (UI) libraries have increasingly dominated the modern internet. However, existing libraries are primarily made for run-time execution, and rely on the user to load and render web applications. This led us to create Million.js, a fast compiler-augmented virtual Document Object Model (DOM) for the web. Million.js reduces load time and time-to-interactive by creating a compiler to compute interactive regions of a web application before the user visits the page. The virtual DOM run-time optimizes interactive content through compiler flags, compute batching, scheduling, and reactive data primitives to achieve optimal performance. When benchmarked against the most popular virtual DOM libraries, Million.js resulted in 133\% to 300\% faster rendering and 2347\% faster load. In a real-world web application with both comparative benchmarks and an informal user study, Million.js loaded 35.11\% faster after migrating from React. The findings show that web applications have the potential to be orders of magnitude faster through JavaScript UI libraries that use Million.js.
\end{abstract}

%
%

\ccsdesc[500]{Human-centered computing~Web-based interaction}
\ccsdesc[500]{Information systems~Web applications}

\keywords{Web technologies, Web Frameworks, JavaScript, User interface (UI), Virtual DOM, Compilers, Optimization}

\maketitle

\input{samplebody-conf}

\bibliographystyle{ACM-Reference-Format}
\bibliography{sample-bibliography} 

\end{document}

%% file: samplebody-conf.tex
\section{Introduction}

\begin{displayquote}
	\textit{My advice for anyone who wants to make a dent in the future of web development: time to learn how compilers work.\\
		\\
		\hspace*{\fill} - Tom Dale \cite{Dale_2017}}
\end{displayquote}

Websites were initially written in Hypertext Markup Language (HTML) and Cascading Style Sheets (CSS) and designed to be static, which limited user interaction \cite{Wieruch_2017}. Eventually, browsers introduced JavaScript and the DOM to provide basic interactivity \cite{MDN_2022, garrett2005ajax}, however, web applications that required significant user interaction to operate suffered scaling and maintenance issues \cite{Zakas_2010}. To support interactive web applications, single-page applications (SPA) were created to allow state-driven development (SDD) through the virtual DOM and its variants \cite{Grov_2015}. However, due to inefficient rendering strategies, SPA and virtual DOM-based web applications have become slower to load and render on user devices \cite{Zakas_2010}, which can lead to worse user retention and experience \cite{Pavic_Anstey_Wagner_2020, Glynn_2020}. 

A wide variety of meta-frameworks (such as Next.js \cite{Vercel_2016} and Astro \cite{AstroTechnologyCompany_2022}) are currently used as a solution to improve the load time of static content in SPA \cite{dinku2022react}. However, these frameworks do not address the virtual DOM's fundamental problem of slow rendering speed on interactive content.

As a result, when web developers want to improve the rendering speed of interactive content, they have to choose between refactoring to a DOM-based implementation or fine-tuning their virtual DOM code with manual optimizations (e.g. memoization \cite{suresh2017compile}, conditional rendering, state management). Choosing the DOM will yield the most efficient results, but worsens the developer experience \cite{Zakas_2010}. Increasing the rendering speed of the virtual DOM requires significant knowledge of internals and may be difficult to optimize to a meaningful degree \cite{kishore2020evolution}.

This paper presents Million.js, a virtual DOM that optimizes both load time and rendering speed. Million.js has a compiler that can be integrated into JavaScript UI libraries and meta-frameworks to optimize interactive content. When benchmarked, Million.js yields 133\% to 300\% faster rendering with the JS Framework Benchmark \cite{Krausest_2015} and 2347\% faster load time than React with the Chrome Devtools Performance benchmark \cite{saverimoutou2018web}. 

Million.js is implemented as an open-source JavaScript library, so web developers can use Million.js wherever they use JavaScript, primarily within browser environments. As such, Million.js is used in 85 open-source projects. In a real-world application using the Million.js React compatibility layer, there was a 35\% improvement in rendering speed after migrating from React.

The paper makes the following contributions: 

\begin{itemize}
	\item \textbf{Million.js}—a compiler-augmented virtual DOM—released as open-source software that can be downloaded and used today\footnote{\url{https://github.com/aidenybai/million}}.
	\item the \textbf{Million.js React compatibility layer}, which allows web developers and existing projects to use React's API.
	\item a \textbf{series of benchmarks and a real-world test} that measure the load time and rendering speed of Million.js.
\end{itemize}

\section{Related Work}

Million.js extends prior work on SDD, SPA, and virtual DOM approaches in JavaScript UI library development.

\subsection{Event-Driven Development}

JavaScript UI library authors have created tools and libraries to simplify the process of creating web applications \cite{zou2014virtual, ladan2015comparing}. One of these libraries is jQuery, a widely used DOM utility library. jQuery's ergonomic API helped developers solve inconsistencies with the verbose DOM syntax across browsers \cite{Muyldermans_2019}. As a result, jQuery flourished in usage in the JavaScript ecosystem \cite{persson2020javascript}. 

Although jQuery had a monumental impact on web development, many web developers faced problems with the imperative event-driven API of jQuery and the DOM \cite{domanski2014javascript}. While jQuery could reasonably handle interactivity for static websites, it became unscalable as web applications became more complex \cite{delcev2018modern, ladan2015comparing}, leading developers to create custom solutions \cite{meyerovich2009flapjax}. 

\subsection{SDD and Virtual DOM}

To address the limitations of the DOM and jQuery, software engineers at Facebook created React, a JavaScript UI library that enabled SDD through the virtual DOM.

\begin{figure} 
	\centering
	\textcolor{lightgray}{\frame{\includegraphics[width=0.8\columnwidth]{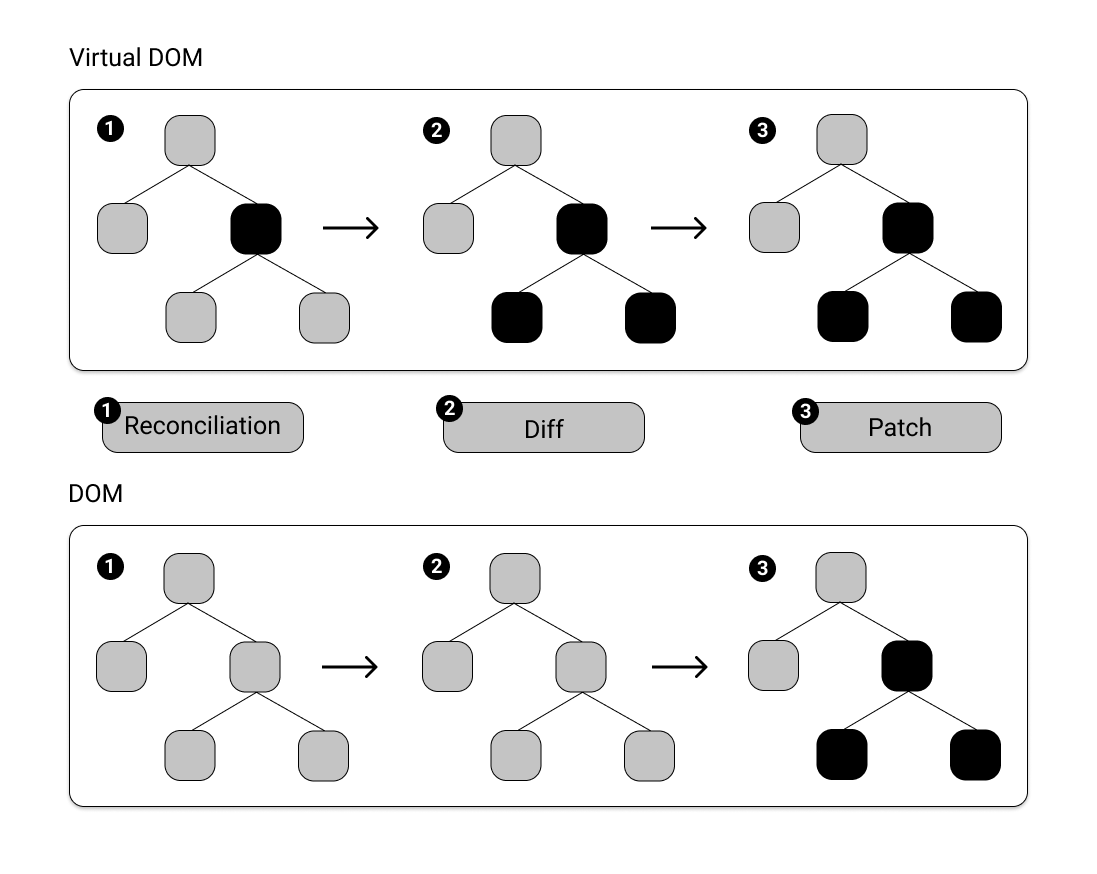}}}
	\caption{Example of virtual DOM diff and patch steps. The reconciliation process includes the diff and patch steps. The diff step calculates the changes to the UI when given the current and new UI. The patch step takes the calculated UI changes from the diff step and uses the DOM to update the UI.}~\label{fig:8-simple-vdom}
\end{figure}

The creation of React and the virtual DOM subsequently enabled state-driven web applications that minimized DOM manipulations, reflows, and repaints \cite{Levlin_2020} without needing to describe control flow. The virtual DOM is a tree diffing algorithm that finds differences between the current and new virtual DOM tree to update UI (see Figure~\ref{fig:8-simple-vdom}).  This led to the implementation of the virtual DOM in other JavaScript UI libraries like Angular, Ember, and Vue \cite{gizas2012comparative}.

Unlike other JavaScript UI libraries, React has its own domain-specific language (DSL) called JSX \cite{gackenheimer2015jsx}, which “provides an extension to JavaScript by adding XML-like syntax that is similar to HTML” \cite{Muyldermans_2019}. React uses the virtual DOM through JSX inside functional components, a superset of JavaScript that compiles down to the virtual DOM, while Vue and Angular use directives and Single-File Components (SFC) to set boundaries and compile to the virtual DOM \cite{kumar2016comparative}. Thus, different API abstraction decisions in libraries allow developers to choose a JavaScript UI library that fits their project \cite{gizas2012comparative} through the virtual DOM at the cost of load time and rendering speed \cite{Levlin_2020}. 

\subsection{Meta-frameworks}

\begin{figure}
	\centering
	\textcolor{lightgray}{\frame{\includegraphics[width=0.8\columnwidth]{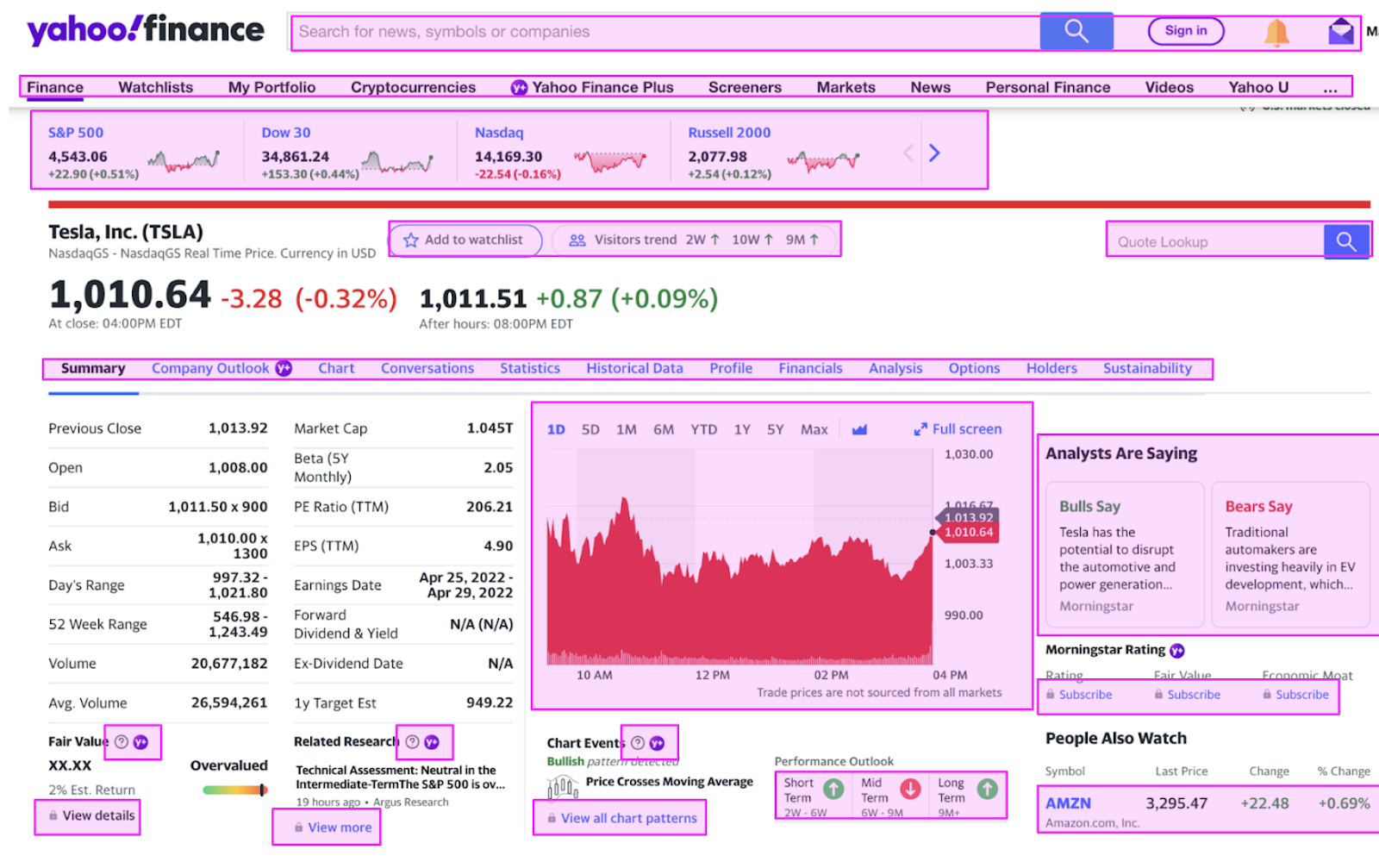}}}
	\caption{Yahoo Finance web application with \textcolor[rgb]{1, 0, 1}{interactive content highlighted in pink}. Static content is loaded before the user visits the page and interactive content will be hydrated upon page visit.}~\label{fig:4-yahoo}
\end{figure}

To improve load time, techniques like Server Side Rendering (SSR) and Static Site Generation (SSG) \cite{petersen2016static} have led to the creation of meta-frameworks like Next.js \cite{Vercel_2016}, which augment and improve the load time of JavaScript UI libraries like React and Vue, using pre-rendering and hydration \cite{iskandar2020comparison}. Pre-rendering turns the web application static for initial load, and hydration turns the static page into an interactive page. Meta-frameworks like Astro \cite{AstroTechnologyCompany_2022} are exploring how SSG and SSR can be optimized with partial hydration, a subset of hydration that only parts of the page interactive (Figure~\ref{fig:4-yahoo}). These techniques can significantly improve the performance of web applications that require less interactivity, such as blogs or eCommerce websites \cite{iskandar2020comparison}.

\subsection{Compiled JavaScript UI Libraries}

While meta-frameworks such as Next.js have helped optimize static content, new JavaScript UI libraries, including Svelte, leverage an optimizing compiler \cite{kennedy2001optimizing} to address the performance constraints of the interactive content of a web application \cite{Harris_2018}. Svelte uses reactive assignments to generate DOM manipulations instead of an intermediate representation of the DOM \cite{Muyldermans_2019}. With this technique, Svelte significantly improves the performance of value-based UI updates but suffers from large structural UI updates where the virtual DOM excels \cite{zou2014virtual}.

With explorations into compiled frameworks, reactive assignments, and efficient hydration, comparative benchmark results of virtual DOM libraries (such as \textbf{\texttt{snabbdom}} \cite{Vindum_2015}, \textbf{\texttt{virtual-dom}} \cite{Esch_2014}, \textbf{\texttt{tiny-vdom}} \cite{Bai_2022}, \textbf{\texttt{simple-virtual-dom}} \cite{Livoras_2015}) are slower (depending on architectural decisions, see Table~\ref{fig:algorithmic-differences}) in contrast to JavaScript UI libraries that utilize these new techniques \cite{mariano2017benchmarking}.

\begin{table*}[]
	{\resizebox{\textwidth}{!}{
\begin{tabular}{@{}llllll@{}}
\toprule
\textbf{Library} & \textbf{Keyed Algorithm}   & \textbf{Virtual Nodes} & \textbf{Render Strategy} & \textbf{Composability} & \textbf{Size}        \\ \midrule
\textbf{\texttt{snabbdom}}         & Longest Common Subsequence & Mutable                & One pass                 & \texttt{node}, \texttt{props}, and \texttt{children} separation   & 4.93kB \\
\textbf{\texttt{virtual-dom}}        & N/A   & Immutable & Two pass & \texttt{diff} and \texttt{patch} separation & 6.03kB \\
\textbf{\texttt{tiny-vdom}}          & Naive & Immutable & One pass & N/A                  & 0.58kB \\
\textbf{\texttt{simple-virtual-dom}} & Naive & Immutable & Two pass & \texttt{diff} and \texttt{patch} separation   & 2.14kB \\ \bottomrule
\end{tabular}
}}
\caption{Differences between methodology and algorithmic differences of the virtual DOM libraries: \texttt{snabbdom}, \texttt{virtual-dom}, \texttt{tiny-vdom}, \texttt{simple-virtual-dom}. The longest common subsequence algorithm is the most efficient keyed algorithm. Mutable virtual nodes are less memory efficient compared to immutable virtual nodes unless used with a one-pass render strategy. More separation means more composability and easier adaptation of the API for different needs.}
\label{fig:algorithmic-differences}
\end{table*}

We extend this prior work by providing the first virtual DOM that can integrate with a compiler.

\section{Overview of Million.js}

Million.js is a compiler-augmented virtual DOM for the web. The software is an open-source JavaScript library and is published on Node Package Manager (NPM) and various Content Delivery Networks (CDN). The tool is built using the TypeScript language and can be used to build new JavaScript UI libraries or modify user interfaces directly 
\cite{greenberg2007declarative}. As such, Million.js is compatible with JavaScript UI libraries that use a virtual DOM, like React.

Million.js contains two key components: the virtual DOM run-time and the optimizing compiler. The virtual DOM run-time updates the UI, similar to existing virtual DOM libraries. However, unlike these libraries, Million.js uses an optimizing compiler to analyze JavaScript and compute unnecessary work during compile time to reduce load time and speed up rendering. Together, the virtual DOM and the optimizing compiler allow for declarative UI development; in particular, it gives authors access to the underlying rendering process to construct customized JavaScript UI libraries \cite{voutilainen2016synchronizing}. 

\begin{table}
	{\resizebox{\columnwidth}{!}{
		\begin{tabular}{@{}llll@{}}
			\toprule
			\textbf{Principle} &
			\textbf{DOM} &
			\cellcolor[HTML]{ea9999}\textbf{Virtual DOM} &
			\cellcolor[HTML]{b7d8a9}\textbf{Million.js} \\ \midrule
			Load time &
			1x &
			\cellcolor[HTML]{ea9999}{\textless{}27.1x slower} &
			\cellcolor[HTML]{b7d8a9}\textless{}1.5x slower \\
			Render speed &
			1x &
			\cellcolor[HTML]{ea9999}1.7x slower &
			\cellcolor[HTML]{b7d8a9}\textless{}1.1x slower \\
			Developer friendly & N/A & $\sim$560 stars/year & 5.4k stars/year \\ \bottomrule
		\end{tabular}
	}}
	\caption{Million.js design principles and Objective and Key Results (OKR) table. All objectives were achieved within one year. Million.js achieved under 1.2x load time and within <1.1x render speed. Load time is measured using Chrome Devtools and render speed with the JS Framework Benchmark.}
	~\label{fig:5-okr}
\end{table}

To optimize for performance and developer usage, Million.js establishes three main design principles in Figure~\ref{fig:5-okr}: fast load time, fast render speed, and ease of use. The virtual DOM suffers from slow load time and rendering speeds, so Million.js aims to improve them by increasing parity with DOM rendering speeds. Each principle is measured, and key results are established to measure the success of Million.js.

\subsection{Implementation}

The virtual DOM render process has three main steps: compile, diff, and patch as seen in Table~\ref{fig:5-okr}. Each phase is completed in different stages and has different purposes. The compile phase is executed once before the user visits the page, and the diff and patch phases are executed at run-time upon rendering.

\begin{figure*} 
	\centering
	\includegraphics[width=0.95\textwidth]{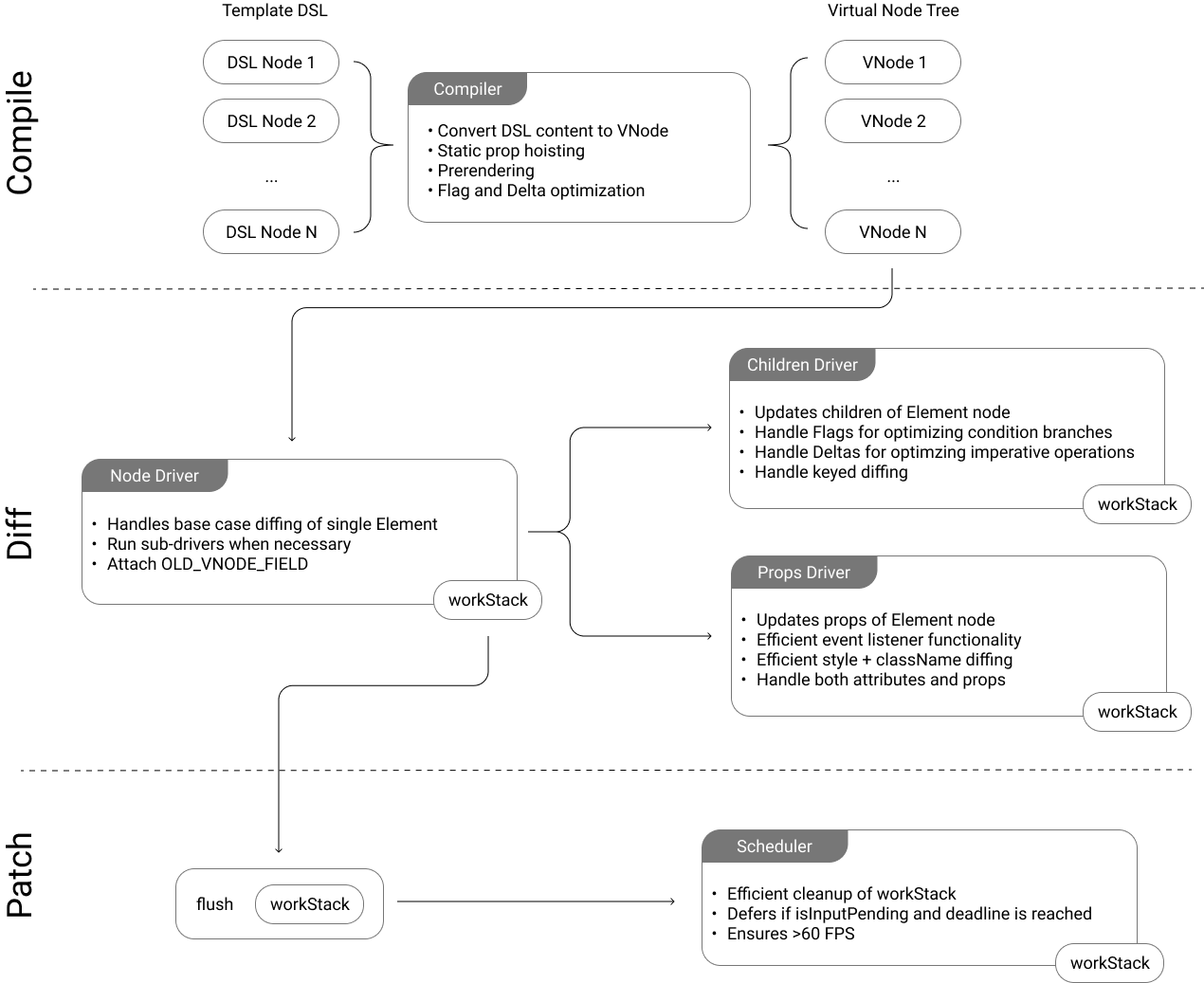}
	\caption{Flowchart of Million.js virtual DOM render process separated into three steps. The compiler step optimizes the interactive content of a website. The diff step is a composable API that is used to calculate the changes to the UI when given the current and new UI. The patch step applies diff changes to the UI.}~\label{fig:6-implementation.png}
\end{figure*}

\subsubsection{Compile Step}

The compile step can occur in conjunction with a meta-framework such as Next.js, or during build time with an SSG tool. Within the compile step, an optimizing compiler will statically analyze the DSL of the source code to find redundant or unnecessary code. While the meta-framework or SSG tool will pre-render the user interface (UI) and hydrate the HTML page, the compiler will optimize the interactive content as seen in Figure~\ref{fig:6-implementation.png}.

The optimizing compiler implements virtual node flattening \cite{keller1998flattening} and static tree hoisting \cite{pawlik2016tree} to improve the load time and rendering speed of interactive content (see Figure~\ref{fig:vnode-code-samples}. It targets the output of DSL compilation: hyperscript, a set of function calls that construct a virtual node tree at run-time \cite{islam2017reactjs}. Hyperscript is used because it is an easy target for DSL compilation but comes at the cost of unnecessary memory allocation during run-time. The virtual node flattening approach addresses this issue by flattening each function call within the hyperscript during the compile step, removing the need to allocate memory during run-time. Within virtual node flattening, compiler flags \cite{ballal2015compiler} are assigned to reduce run-time computation (similar to Inferno \cite{Inferno_2020}). Along with virtual node flattening, the static tree hoisting approach further reduces unnecessary memory allocation by moving flattened hyperscript that does not depend on the state to the global JavaScript scope.

\begin{figure}
	\centering
	
    \begin{verbatim}

h('div', null, 'Hello World');
// Turns into:
{ 
    tag: 'div', 
    children: ['Hello World'], 
    flag: Flags.ONLY_TEXT_CHILDREN 
};
    \end{verbatim}
	\caption{Code samples of hyperscript and flattened and hoisted virtual node, ordered by less to more optimizations by the compiler. The latter code sample is a compiler optimization exclusive to Million.js.}~\label{fig:vnode-code-samples}
\end{figure}

\subsubsection{Diff Step}

The diff step occurs on the user’s device during run-time upon each render to calculate the changes to the UI when given the current and new UI. In Figure~\ref{fig:8-simple-vdom}., the virtual DOM calculates the specific changes during the diff step and applies those changes to the UI during the patch step.

\begin{figure}
	\centering
	\textcolor{lightgray}{\frame{\includegraphics[width=0.9\columnwidth]{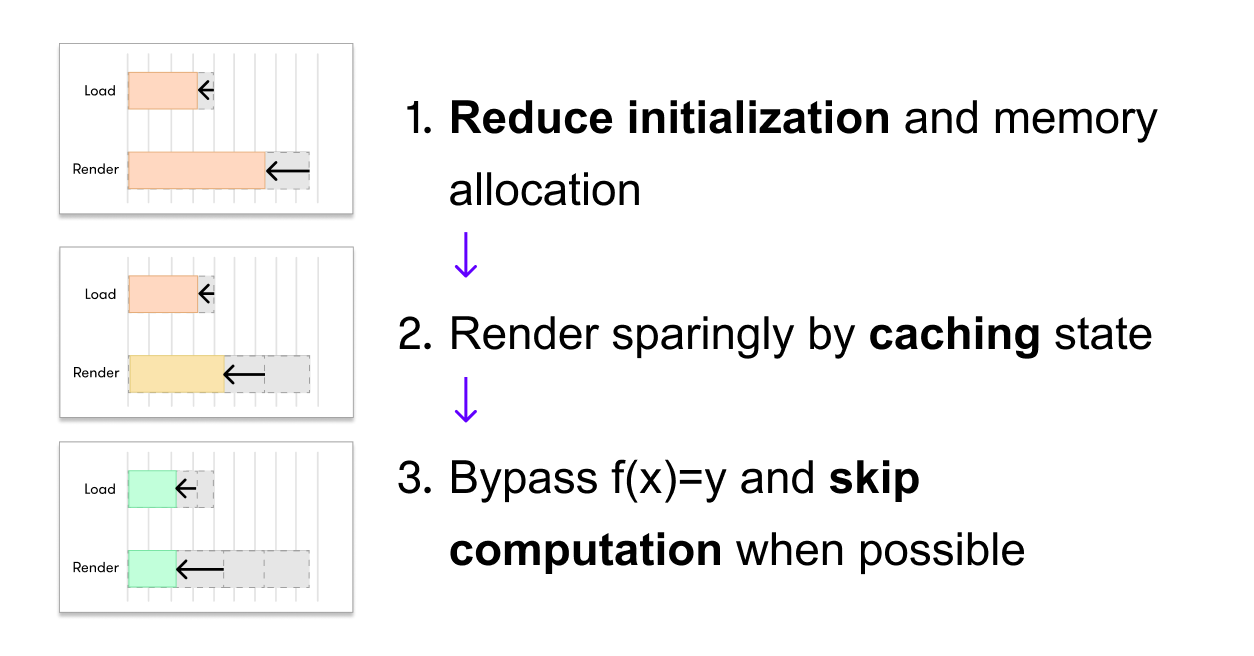}}}
	\caption{Example of fast path optimization represented as a to-do list. Step 1 occurs during the compile step, step 2 occurs during the diff step, and step 3 occurs during the patch step. Every step works together to optimize the rendering speed and load time of UI updates.}~\label{fig:9-todo}
\end{figure}

Diff and patch steps exist in some capacity within all virtual DOM implementations. Million.js uses a two-pass, immutable virtual node run-time, with separation at the \texttt{node}, \texttt{props}, and \text{children} level (see Table~\ref{fig:algorithmic-differences} for comparison). The main focus, however, is improvement in the performance of the diff and patch steps by integrating fast paths for the compile step to reduce extraneous diffing. When fast paths are used during run-time, the virtual DOM allocates less memory and performs less computation, leading to faster load times and rendering speeds \cite{golestani2019characterization}. These fast paths are keyed nodes \cite{antonio2015inside}, compiler flags, and delta. The methodology can be visualized in Figure~\ref{fig:9-todo}.

The keyed nodes approach is a run-time optimization that reduces the amount of computation needed by identifying virtual children nodes with a key. Each key represents a unique virtual node, meaning nodes can be moved instead of basic update and delete operations. Moving keyed nodes is a technique used in a special algorithm to reduce computation based on an \texttt{O(ND)} algorithm \cite{apostolico1986improving, myers1986ano}. By removing the need to diff head and tail nodes, a key map can be generated to efficiently move nodes to reduce computation when changes between virtual children nodes are minimal.

Compiler flags are generated during the compile step to change the behavior of the diff step. They are used to infer the shape of virtual nodes during run-time to skip conditions of the diff step or invoke more efficient algorithms. By default, virtual nodes have the same flag, but when virtual children node edge cases are met, such as no children, only text children, or only keyed children, less computation is used to diff the virtual node.

Delta is special data that can be attached to virtual nodes that provide an escape hatch for imperative UI changes. Delta can be used by the compile step to skip the diff step entirely, or in run-time by the web developer. This is useful for feed-like UI, such as post feeds or to-do lists.

Along with the aforementioned fast paths, Million.js amortizes computation with strategies like scheduling and batching. Traditionally, virtual DOM implementations use a single, uninterrupted, synchronous render. Once an update starts rendering, nothing can interrupt it until the user can see the result on the screen. This includes user interactions, meaning that the user cannot interact with the page until rendering is complete. Million.js introduces scheduling, which prioritizes more urgent CPU-bound updates (such as creating DOM nodes and running component code \cite{degenbaev2016idle}) and defers less important updates to user interactions. Batching is also used to cancel intermediate renders when similar updates are executed in succession and should be used in conjunction with scheduling \cite{wloka2003batch}.

\subsubsection{Patch step}

The patch step takes the calculated UI changes from the diff step and applies those changes to the UI. The diff and patch steps run in a two-pass format to separate virtual DOM change calculation with browser reflow and repaint calculations \cite{ast2016incremental}. Separating these two categories allows the browser to optimize calculations for both run-time steps \cite{Kaul_2020}.

\subsection{React Compatibility Layer}

Million.js provides a fast underlying virtual DOM to update user interfaces. However, the virtual DOM is not intended to be directly used by web developers, as it does not provide concepts like state or components that define JavaScript UI libraries. Thus, we created a React compatibility layer that allows web developers to start existing or migrate React projects to Million.js, without having to re-learn a new API.

The React compatibility layer integrates with existing React projects like Vite \cite{Vite_2022} and meta-frameworks like Next.js through a plugin. The plugin injects a run-time that imitates the React API implemented with Million.js internals. Some limitations of the API include new React version 18 concurrent mode features like Suspense, HTML streaming, and React Server Components \cite{React_2022}. As such, the React compatibility layer lacks the ability to prioritize renders out of order, a bonus feature for increasing frame rate.

One notable feature Million.js provides through the React compatibility layer a reactive data primitive that performs at optimal speeds with constant time complexity by bypassing the diff step altogether. Feeds or list-like UI are prevalent in modern web applications and have the potential for optimization. With traditional virtual DOM, lists have linear time complexity, meaning rendering speed and memory allocation gets worse as more entries are loaded \cite{dermawan2020encrypted}. Through the Delta optimization, direct imperative UI changes can be made using the JavaScript Proxy API \cite{keil2015transparent}, greatly reducing the computation and increasing rendering speed.

\section{Evaluation}

Three benchmarks are used to determine the performance of Million.js: a re-implementation of a subset of the JS Framework Benchmark \cite{Krausest_2015} measuring rendering speed, a Devtools performance benchmark measuring load time, and real-world application performance. These benchmarks measure render speed in operations per second and load time in milliseconds to quantify the performance of rendering methods \cite{ratanaworabhan2010jsmeter}. 

\subsection{JS Framework Benchmark}

We tested ten different measurement benchmark suites on seven different implementations of web page rendering using a subset of the JS Framework Benchmark, as seen in Table~\ref{fig:11-render-res}. The first type of method is the raw rendering method: \textbf{DOM}, \textbf{\texttt{innerHTML}}. The second type of method is the virtual DOM library: \textbf{Million.js}, \textbf{\texttt{snabbdom}} \cite{Vindum_2015}, \textbf{\texttt{virtual-dom}} \cite{Esch_2014}, \textbf{\texttt{tiny-vdom}} \cite{Bai_2022}, \textbf{\texttt{simple-virtual-dom}} \cite{Livoras_2015}.

The virtual DOM libraries, except for Million.js and \texttt{snabbdom}, had fewer operations per second when compared with raw DOM rendering implementations, with the worst performing implementations being \texttt{tiny-vdom} and \texttt{simple-virtual-dom}.

Raw rendering implementations, like DOM and \texttt{innerHTML} had relatively higher operations per second in aggregate insertions or deletions, but relatively lower operations per second in partial or complex updates when compared to virtual DOM implementations. Additionally, implementations with keyed diffing, specifically Million.js and \texttt{snabbdom}, had higher operations per second when compared to the other virtual DOM libraries.

\begin{table*}
	{\resizebox{\textwidth}{!}{%
		\begin{tabular}{@{}
			>{\columncolor[HTML]{FFFFFF}}l 
			>{\columncolor[HTML]{B6D7A8}}c cc
			>{\columncolor[HTML]{FFF2CC}}c cc
			>{\columncolor[HTML]{EA9999}}c @{}}
			\toprule
			\textbf{Method (ops/s)}                                     & \cellcolor[HTML]{FFFFFF}\textbf{DOM}   & \cellcolor[HTML]{FFFFFF}\textbf{million} & \cellcolor[HTML]{FFFFFF}\textbf{innerHTML} & \cellcolor[HTML]{FFFFFF}\textbf{snabbdom} & \cellcolor[HTML]{FFFFFF}\textbf{virtual-dom} & \cellcolor[HTML]{FFFFFF}\textbf{tiny-vdom} & \cellcolor[HTML]{FFFFFF}\textbf{simple-virtual-dom} \\ \midrule
			Append 1,000 rows to a table of 10,000 rows & \cellcolor[HTML]{B7D8A9}\textbf{21.7} ±0.9\%  & \cellcolor[HTML]{B6D7A8}\textbf{21.8} ±3.3\%    & \cellcolor[HTML]{F2EEC6}\textbf{9.4} ±2.9\%       & \textbf{6.6} ±5.9\%                              & \cellcolor[HTML]{F7D3BA}\textbf{5.9} ±3.5\%         & \cellcolor[HTML]{EFB1A6}\textbf{5.1} ±3.5\%       & \textbf{4.6} ±3.1\%                                        \\
			Clear a table with 1,000 rows               & \textbf{207} ±2.4\%                         & \cellcolor[HTML]{C6DDB0}\textbf{190} ±2.4\%   & \cellcolor[HTML]{D7E3B8}\textbf{172} ±2.9\%     & \textbf{128} ±4.3\%                            & \cellcolor[HTML]{F5C7B3}\textbf{97.7} ±3.8\%        & \cellcolor[HTML]{F3C2B0}\textbf{93.7} ±3.6\%      & \textbf{64.1} ±2.3\%                                       \\
			Create 10,000 rows                          & \textbf{36.8} ±4.6\%                          & \cellcolor[HTML]{D6E3B8}\textbf{30.5} ±5\%    & \cellcolor[HTML]{BAD9AA}\textbf{36.1} ±4.8\%       & \cellcolor[HTML]{FEF1CB}\textbf{22.2} ±1.9\%     & \cellcolor[HTML]{FFF2CC}\textbf{22.3} ±1.2\%        & \cellcolor[HTML]{EFB1A7}\textbf{12.9} ±3.4\%    & \textbf{9.3} ±4.6\%                                        \\
			Create 1,000 rows                           & \textbf{36} ±5.2\%                          & \cellcolor[HTML]{D5E3B7}\textbf{30.3} ±4.7\%    & \cellcolor[HTML]{B7D8A9}\textbf{35.8} ±4.6\%      & \cellcolor[HTML]{FEF1CB}\textbf{22.2} ±2.1\%     & \cellcolor[HTML]{FFF2CC}\textbf{22.4} ±1.6\%        & \cellcolor[HTML]{EFB1A7}\textbf{12.7} ±3.6\%      & \textbf{9.0} ±4.9\%                                        \\
			Update every 10th row for 1,000 rows        & \cellcolor[HTML]{B7D8A9}\textbf{220} ±3.6\% & \cellcolor[HTML]{B6D7A8}\textbf{221} ±3.3\%   & \cellcolor[HTML]{E2E8BE}\textbf{135} ±3.6\%     & \textbf{76.4} ±5.4\%                             & \cellcolor[HTML]{F9DABE}\textbf{69.8} ±3.4\%        & \cellcolor[HTML]{F0B3A8}\textbf{58.7} ±4\%      & \textbf{51.2} ±3.7\%                                       \\
			Remove a random row from 1,000 rows         & \textbf{225} ±5.2\%                         & \cellcolor[HTML]{C2DCAE}\textbf{202} ±3.8\%   & \cellcolor[HTML]{E1E7BD}\textbf{138} ±3.7\%     & \textbf{74.2} ±8.9\%                             & \cellcolor[HTML]{F6CFB8}\textbf{57.6} ±6.7\%        & \cellcolor[HTML]{EFAEA5}\textbf{41.7} ±7.8\%      & \textbf{31.4} ±5.1\%                                       \\
			Update 1000 rows                            & \cellcolor[HTML]{BFDBAD}\textbf{119} ±1.9\% & \cellcolor[HTML]{B6D7A8}\textbf{127} ±3.3\%   & \cellcolor[HTML]{BBD9AB}\textbf{123} ±2.8\%     & \cellcolor[HTML]{FADEC0}\textbf{50.7} ±0.4\%     & \cellcolor[HTML]{FFF2CC}\textbf{55.5} ±6.7\%        & \cellcolor[HTML]{F2BDAD}\textbf{42.3} ±5.2\%      & \textbf{33.2} ±6.1\%                                       \\
			Select a random row from 1,000 rows         & \textbf{210} ±5.9\%                         & \cellcolor[HTML]{B8D8A9}\textbf{208} ±5.7\%   & \cellcolor[HTML]{E5E9BF}\textbf{123} ±5.3\%     & \textbf{72.5} ±7.5\%                             & \cellcolor[HTML]{FADEC0}\textbf{66.6} ±7.2\%        & \cellcolor[HTML]{EB9D9B}\textbf{46.8} ±6.3\%      & \textbf{45.4} ±4.9\%                                       \\
			Swap 2 rows for table with 1,000 rows       & \textbf{224} ±3.5\%                         & \cellcolor[HTML]{C3DCAF}\textbf{198} ±4.4\%   & \cellcolor[HTML]{EAEBC2}\textbf{118} ±5.6\%     & \textbf{73.5} ±8.2\%                             & \cellcolor[HTML]{F7D0B8}\textbf{57.9} ±8.9\%        & \cellcolor[HTML]{F1B9AB}\textbf{46.8} ±6.3\%      & \textbf{31.6} ±5.8\%                                       \\
			Geometric mean                              & \textbf{105.7} ±3.5\%                         & \cellcolor[HTML]{BFDBAD}\textbf{98.8} ±3.9\%    & \cellcolor[HTML]{DCE5BB}\textbf{74.3} ±3.9\%      & \textbf{44.1} ±3.7\%                             & \cellcolor[HTML]{FADFC1}\textbf{39.8} ±4\%        & \cellcolor[HTML]{F0B4A8}\textbf{29.7} ±4.7\%      & \textbf{23.1} ±3.4\%                                       \\ \bottomrule
		\end{tabular}%
	}}
	\caption{JS Framework Benchmark compatible benchmarks tested on Chrome/94.0.4606.71. \textcolor{benchGreen}{Green colors} signify more operations per second, and \textcolor{benchRed}{red colors} signify fewer operations per second. Million.js ranked the second-best scores overall in the benchmark of seven implementations with the geometric mean of benchmarks at 98.8 operations per second.}
	\label{fig:11-render-res}
\end{table*}

\begin{figure}
	\centering
	\textcolor{lightgray}{\frame{\includegraphics[width=0.99\columnwidth]{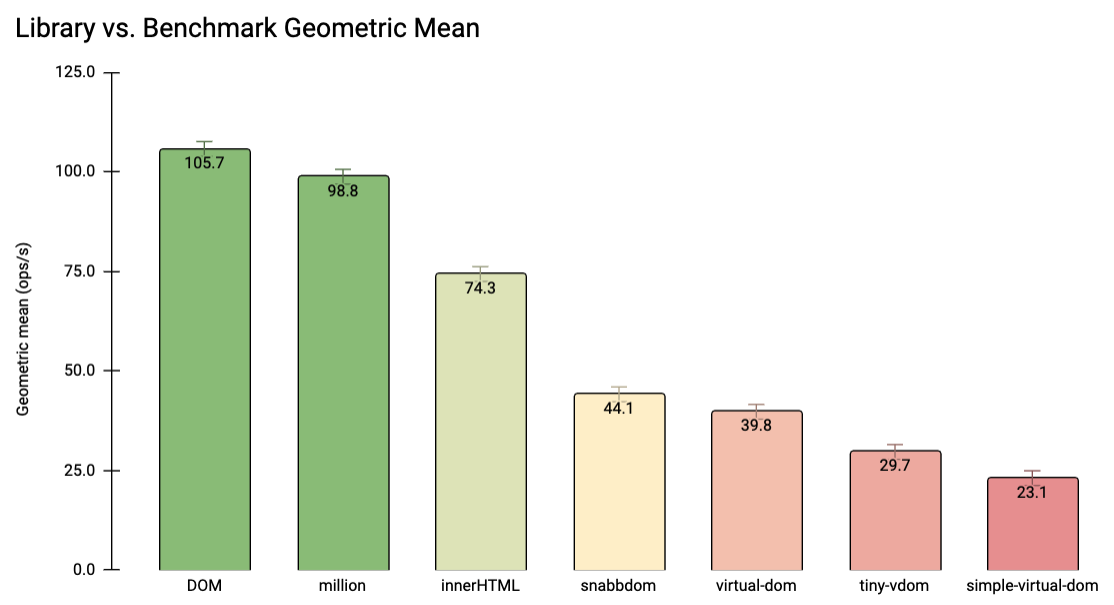}}}
	\caption{Geometric means of render speeds for benchmarks measured in operations per second. \textcolor{benchGreen}{Green colors} signify more operations per second, and \textcolor{benchRed}{red colors} signify fewer operations per second (higher is better). Million.js ranked the second best scores overall in the benchmark of seven implementations, ranging from 133\% to 300\% faster-rendering speed.}~\label{fig:13-render-graph}
\end{figure}

The geometric mean of benchmark results is visualized in Figured~\ref{fig:13-render-graph}. Additionally, due to uncertainties with benchmark measurements, an average deviation of 1.856 operations per second is also visualized. The DOM method has the highest operations per second of the implementations tested, with a geometric mean of 105.7 operations per second. Million.js is relatively comparable to the DOM method and significantly outperforms the latter implementations by 133\% to 300\%.

Additionally, Million.js has a much smaller bundle size at 1.1 kilobytes compared to \texttt{snabbdom} and \texttt{virtual-dom}, but slightly larger than \texttt{tiny-vdom} at 0.58 kilobytes (see Figure~\ref{fig:algorithmic-differences}). Since Million.js has a composable API, the bundle size can become even lower through tree shaking and dead code elimination \cite{obbink2018extensible}.

\subsection{Devtools Performance Flamegraph}

We used the Chrome Devtools Performance benchmark \cite{saverimoutou2018web} to measure the load time of common interactive UI-like feeds \footnote{\url{https://github.com/aidenybai/million-delta-react-test}}. The benchmark tests the total scripting time of different implementations to add 5000 DOM nodes to the UI consecutively with one iteration after a warm-up round. We tested four different implementations: \textbf{Direct DOM updates}, \textbf{Delta rendering}, \textbf{Keyed rendering}, \textbf{Virtual DOM rendering}. The latter three implementations use the React compatibility layer for Million.js, with Delta rendering using reactive data primitives, keyed rendering using the keyed nodes optimization, and virtual DOM being the default method of rendering Million.js provides.

\begin{figure}
	\centering
	\textcolor{lightgray}{\frame{\includegraphics[width=0.9\columnwidth]{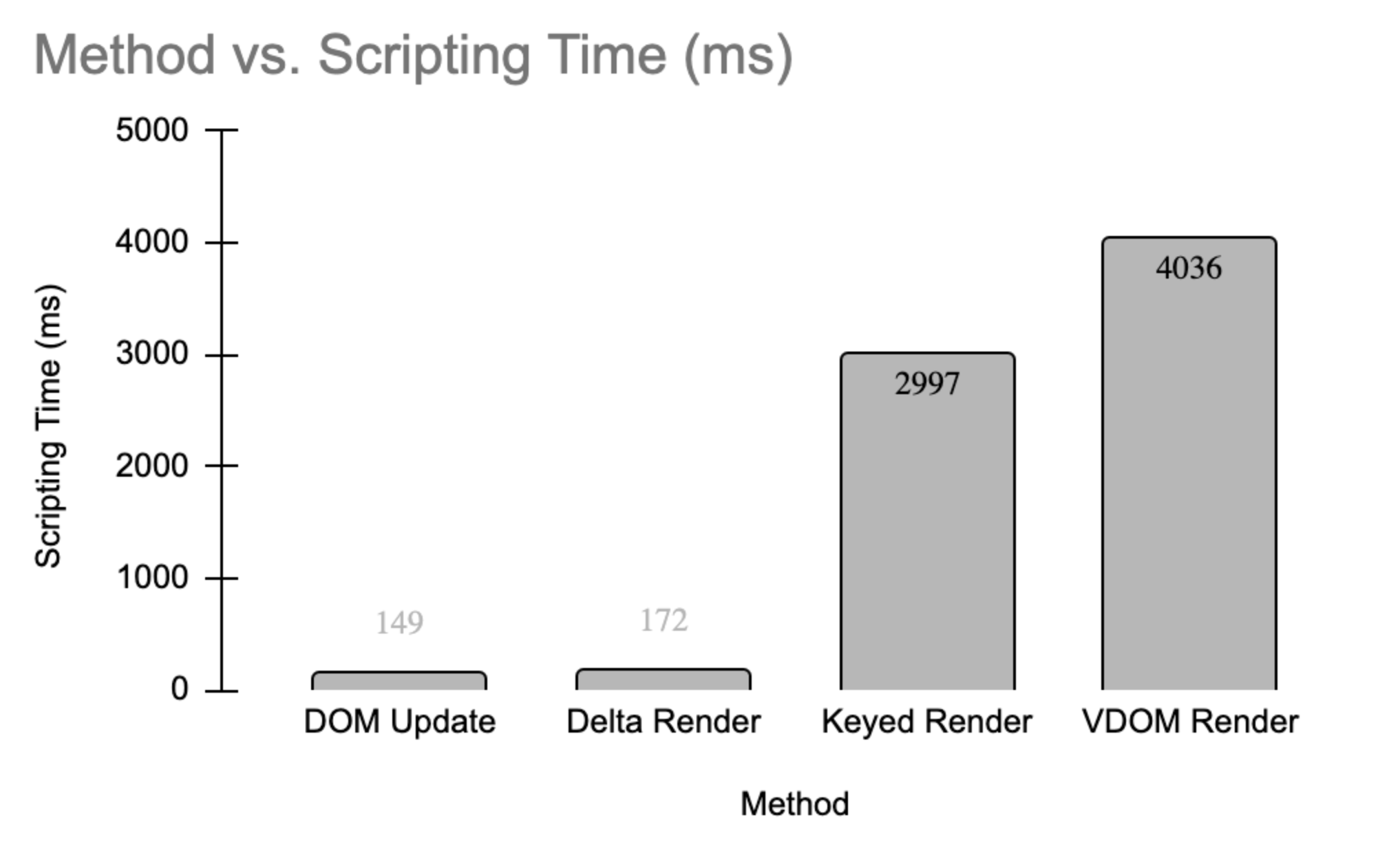}}}
	\caption{Devtools performance results of node append benchmark of DOM update, Delta render, keyed render, and virtual DOM (VDOM) render tested on Chrome/94.0.4606.71. Benchmark for DOM update finished in 149 milliseconds, Delta render in 172 milliseconds, 2997 milliseconds, VDOM render in 4036 milliseconds.}~\label{fig:14-load}
\end{figure}

As seen in Figure~\ref{fig:14-load}, direct DOM updates yielded the lowest scripting time at 149 milliseconds, with a small and consistent call stack. Delta rendering yielded a scripting time of 172 milliseconds and is 15\% longer compared to direct DOM updates. Delta rendering showed characteristics of batching, with some scripting occurring before and after delta calculation shown in green and pink, which is what allows for such low overhead compared to direct DOM. The keyed rendering yielded a scripting time of 4022 milliseconds, with a large call stack due to diff computations, and is 2238\% longer than direct DOM updates. The virtual DOM rendering yielded 4036 milliseconds, with slightly more memory allocation than keyed rendering, and is 2247\% longer than direct DOM updates.

While direct DOM updates yielded the lowest scripting time of all implementations tested, Delta rendering was a fast implementation that used the React compatibility layer, being 24 times faster than virtual DOM rendering.

\subsection{Real World Application Performance}

To assess real-world performance, we integrated the React compatibility layer into the Wyze Web View \cite{Wyze_2022} web application. The Wyze Web View runs on Next.js and displays a live video feed of smart home cameras. The Wyze Web View regularly receives thousands of unique sessions daily, and as a result, is a critical component in the Wyze smart home ecosystem. One of the main issues reported by many Wyze users was that the camera feed would take upwards of several minutes to load cameras completely.

\begin{figure}
	\centering
	\textcolor{lightgray}{\frame{\includegraphics[width=0.99\columnwidth]{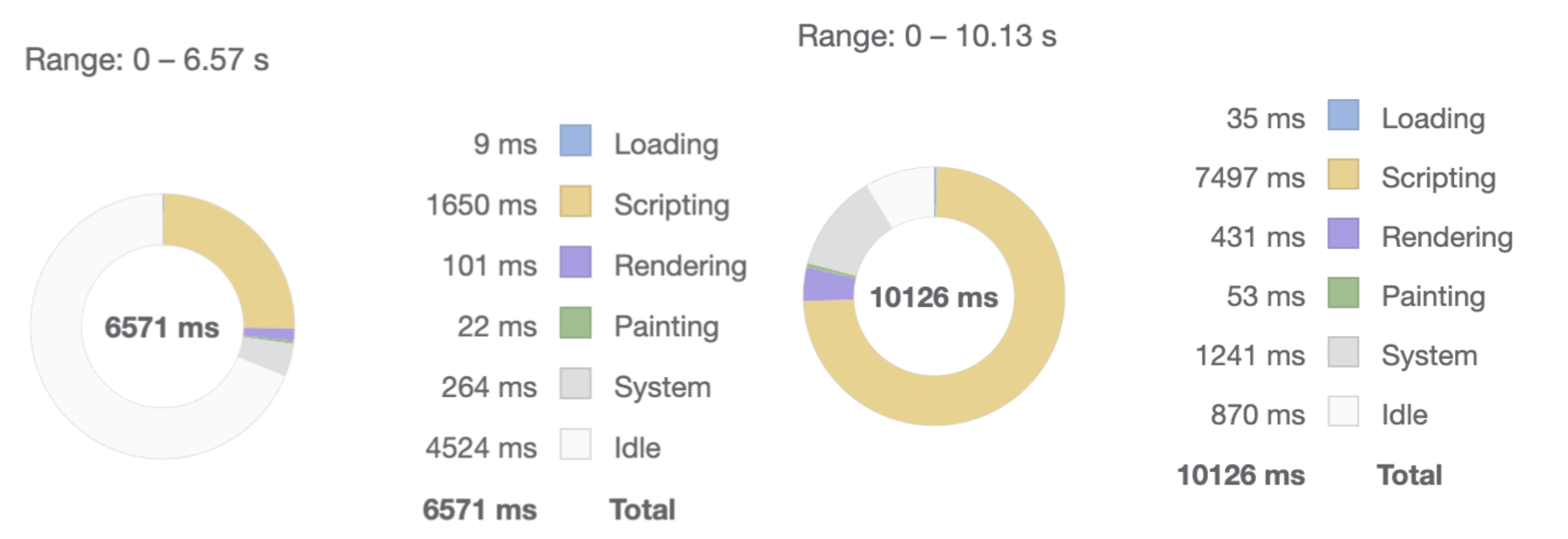}}}
	\caption{Wyze Web View compute time breakdown, left is Million.js’ React compatibility layer and right is React.  The React-based implementation experienced a total compute time of 10126 milliseconds and the new Million.js implementation experienced a total compute time of 6571 milliseconds. Switching to Million.js resulted in a 35.11\% faster load time.}~\label{fig:19-wyze-rendering}
\end{figure}

By migrating the Next.js project from React to Million.js, the Wyze Web View experienced a 35.11\% decrease in compute time taken to render all five cameras as seen in Figure~\ref{fig:19-wyze-rendering}. The React-based implementation experienced a total compute time of 10126 milliseconds and a scripting time of 7497 milliseconds, and the new Million.js implementation experienced a total compute time of 6571 milliseconds, with a scripting time of 1650 milliseconds.

After migrating to the new Million.js implementation, an informal user study was conducted on ten active Wyze Web View users. All reported that they experienced a reduction in load time and less perceived lag with the new improvements.


\section{Discussion \& Future Work}

While the JS Framework Benchmark and the Devtools performance benchmark both are good indicators of rendering speed and load time for large page navigations \cite{mohammadiperformance}, the real-world application performance benchmark is important for assessing the viability of Million.js in a product application scenario. 

In real-world web applications, it is the user code that often is the bottleneck to rendering \cite{maragatham2018study, hardy2021performance, souders2008high}. For instance, the Wyze Web View uses much of the load time to establish a connection to create video feeds. Furthermore, the real-world application benchmark and the Devtools performance benchmark primarily measure load time and don’t necessarily account for subsequent renderings, such as page navigations. Despite this, well-maintained and standardized benchmarks, such as the ones presented in this paper, can lead researchers to learn a lot about how Million.js behaves.

There is still much work to be done in this area. Million.js is open source and other researchers may use Million.js as a starting point for further extensions like new JavaScript UI libraries or high-performance web applications. In the future, more effort will go towards exploring more complex optimization techniques during compilation like static analysis to further improve load time and render speed for UI, increase compatibility with the React ecosystem, and integrate Million.js into more real-world web applications.

\section{Conclusion}

We presented Million.js, a fast compiler-augmented virtual DOM for the web. The initial objectives (see Table~\ref{fig:5-okr}) of Million.js were to create a novel virtual DOM implementation with quick load time, fast render speed, and to make it easy for developers to use.

By leveraging an optimizing compiler and reactive data primitives to improve the load time and render speed of UI, we were successfully able to achieve fast load time and fast render speed. We conducted two benchmarks to measure load time and render speed, showing that Million.js loads 23.47 times faster compared to virtual DOM and renders 133\% to 300\% faster than existing virtual DOM implementations. 

\begin{figure}
	\centering
	\textcolor{lightgray}{\frame{\includegraphics[width=0.99\columnwidth]{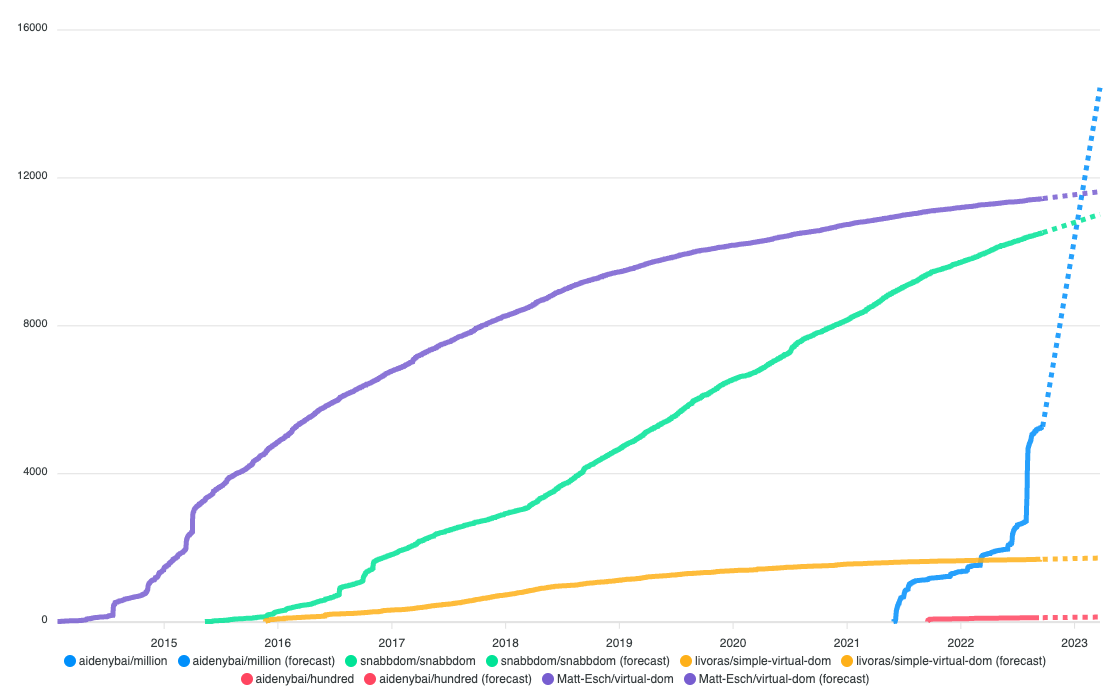}}}
	\caption{Star history graph of \textcolor{stargraphMillion}{Million.js}, \textcolor{stargraphSnabbdom}{\texttt{snabbdom}}, \textcolor{stargraphSimpleVirtualDom}{\texttt{simple-virtual-dom}}, \textcolor{stargraphHundred}{\texttt{hundred}} (previously \texttt{tiny-vdom}), \textcolor{stargraphVirtualDom}{\texttt{virtual-dom}}. Million.js ranked as the third most-starred virtual DOM library and is expected to be the most-starred library in one year.}~\label{fig:20-star-track}
\end{figure}

We were also successfully able to increase developer use. Within one year, Million.js received over 5.1 thousand stars, becoming the third most starred\footnote{\url{https://seladb.github.io/StarTrack-js/}} virtual DOM library on GitHub (see Figure~\ref{fig:20-star-track}), and was featured in major developer publications like JavaScript Weekly, React Status, and React Newsletter. Furthermore, Million.js is actively used in 80 open-source projects on GitHub. The GitHub repository culminated in 1.5 thousand contributions, 166 pull requests, and 70 feature requests.

In the next decade, web users will continue to see improvement in load and rendering speeds for better user retention and experience \cite{Pavic_Anstey_Wagner_2020, GoogleDevelopers_2008, Glynn_2020}. With internet usage growing rapidly in developing nations \cite{wang2011web} and through the  Internet of Things \cite{sin2016performance} market, fast UI performance has never been more important.

\section{Using Million.js}

Million.js is an open-source JavaScript package that users can download using NPM, a Node.js package manager. Million.js can be used in any JavaScript run-time with the JavaScript DOM API. Information on how to download Million.js and the source code can be accessed at \url{https://millionjs.org}.

\section{Acknowledgments}

The work was supported in part by CHS MST Magnet program and Wyze Labs, Inc. We thank Matthew Conlen, the Million.js open-source contributors, and user study participants for their help and valuable input.